\documentclass[twocolumn]{revtex4}

\usepackage{graphicx}
\usepackage{dcolumn}
\usepackage{bm}
\usepackage{mathrsfs}
\usepackage{comment}
\usepackage[usenames,dvipsnames]{color}
\usepackage{xcolor}
\usepackage{slashed}
\usepackage{cancel}
\usepackage{simplewick}
\usepackage{appendix}
\usepackage[colorlinks, linkcolor=red,anchorcolor=green,citecolor=blue]{hyperref}

\graphicspath{{./figs/}}
\begin{document}
\title{Heavy-quark potential in Gribov-Zwanziger approach\\ around deconfinement phase transition}
\author{Wan Wu, Guojun Huang, Jiaxing Zhao and Pengfei Zhuang}
\affiliation{Physics Department, Tsinghua University, Beijing 100084, China}
\date{\today}
\begin{abstract}
The interaction potential between a pair of heavy quarks is calculated with resummed perturbation method in Gribov-Zwanziger approach at finite temperature. The resummed loop correction makes the potential complex. While the real part is, as expected, screened and becomes short-ranged in hot medium, the strength of the imaginary part increases with temperature and is comparable with the real part, which is very different from the previous calculation in HTL approach. This means that, both the color screening and Landau damping play important role in the dissociation of heavy flavor hadrons in hot medium.
\end{abstract}

\maketitle

Considering the large mass $m$ and small velocity $v$, there exists an hierarchy of energy scales $m\gg mv\gg mv^2$ for heavy quarks. When integrating out the momentum larger than $m$ and $mv$ from the Quantum Chromodynamics (QCD) respectively, one obtains the non-relativistic QCD (NRQCD) and potential non-relativistic QCD (pNRQCD) theories~\cite{Caswell:1985ui,Brambilla:1999xf,Brambilla:2004jw} for the study of heavy quark systems. At the leading order, the equation of motion in pNRQCD returns to the Schr\"odinger equation in quantum mechanics, and the dynamics is fully described by an interaction potential. The two-body Schr\"odinger equation with the Cornell potential between a pair of heavy quarks successfully describes the quarkonium properties in vacuum~\cite{Zhao:2020jqu}. The one-gluon exchange between two heavy quarks gives rise to the Coulomb part of the potential, while the confinement part should come from non-perturbative calculations through for instance lattice simulations~\cite{Koma:2006si,Kawanai:2011jt}.

In recent years heavy-flavor hadrons are widely considered as a probe of the new state of matter - quark-gluon plasma (QGP) created in relativistic heavy-ion collisions~\cite{Matsui:1986dk,Zhao:2020jqu,Dong:2019unq,Rothkopf:2019ipj,Chapon:2020heu,Zhao:2020nwy,Zhao:2016ccp,He:2014tga,ExHIC:2017smd}. This extends the study of heavy-quark potential from vacuum to finite temperature. In a hot medium of light quarks and gluons, the heavy-quark potential is expected to become complex: the color screening by the surrounding quarks and gluons reduces the Cornell potential~\cite{Nadkarni:1986as,Wong:2004zr,Satz:2005hx,Kaczmarek:2004gv}, analogy to the Debye screening in electromagnetical systems, and the imaginary part is introduced by the Landau damping or color singlet to octet transition~\cite{Brambilla:2008cx}. At extremely high temperature, the Landau damping becomes dominant, and the potential can be well described by Hard-Thermal Loop (HTL) resummed perturbation~\cite{Laine:2006ns,Beraudo:2007ky}. At finite temperature, the heavy-quark potential can be extracted from the quarkonium spectral functions via lattice QCD simulations~\cite{Rothkopf:2011db,Bazavov:2014kva,Burnier:2014ssa,Burnier:2015tda,Lafferty:2019jpr,Bala:2019cqu}. Recently, the machine learning method~\cite{Shi:2021qri} and spectral extraction strategy~\cite{Larsen:2019zqv,Bala:2021fkm} are used to calculate the heavy-quark potential, both indicate a much larger imaginary part compared with the previous perturbative calculations~\cite{Laine:2006ns,Beraudo:2007ky}.

Since Faddeev and Popov quantized the Yang-Mills theory by introducing a new kind of particles called ghost through the introduction of a gauge~\cite{Faddeev:1967fc}, the quantization has become a standard textbook item. Some years later, however, Gribov discovered that the gauge fixing in the quantization is not complete, there exist still gauge copies called Gribov copies which could affect the infrared region of those gauge quantities such as the gluon and ghost propagators~\cite{Gribov:1977wm}. The Gribov action arises from the restriction of the domain of the integration in Euclidean space to the Gribov region $\Omega$, which is defined as the set of all gauge field configurations fulfilling a gauge (for instance the Landau gauge $\partial^\mu A_\mu^a=0$) and for which the Faddeev-Popov operator $M^{ab}=-\partial_\mu(\delta^{ab}\partial^\mu-gf^{abc}A^\mu_c)$ is strictly positive~\cite{Gribov:1977wm,Dokshitzer:2004ie}. The original Gribov Lagrangian includes a nonlinear term and hard to be calculated. Zwanziger used the BRST method~\cite{Zwanziger:1982na,Zwanziger:1989mf,Zwanziger:1992qr} to modify the Lagrangian through the introduction of a set of auxiliary fields and derived the Gribov-Zwanziger (GZ) Lagrangian in local form which is now used widely~\cite{Zwanziger:2006sc,Cucchieri:2000hv,Gracey:2009mj,Canfora:2015yia}. From the GZ Lagrangian one can conveniently read off the gluon propagator. For instance, it reads in Landau gauge~\cite{Gribov:1977wm} (in Coulomb gauge see Ref.\cite{Burgio:2008jr}),
\begin{equation}
\label{GZ1}
D_{\mu\nu}^{ab}(p)=\delta^{ab}\left(\delta_{\mu\nu}-{p_\mu p_\nu\over p^2}\right){p^2\over p^4+m_G^4},
\end{equation}
where $m_G$ is the Gribov mass parameter. In comparison with the normal gluon propagator in the limit of $m_G\to 0$, the GZ propagator is suppressed in the infrared region by the complex poles at $p^2=\pm im_G^2$. This structure does not allow us to attach the usual particle meaning to the gluon propagator, invalidating the interpretation of gluons as excitation of the physical spectrum. Gluons are confined by the Gribov condition. This means that, the GZ approach successfully describes the confinement of gluons by the non-physical singularities in the gluon propagator.

Considering the fact that the confinement is already reflected in the gluon propagator (\ref{GZ1}) at lowest order, the GZ approach provides a possibility to perturbatively calculate the heavy-quark potential. This can provide a way to understand the physics of the complex potential extracted from the lattice data. In fact, the calculation at one-loop level in vacuum shows a Coulomb potential and a linear term modified by a logarithm~\cite{Golterman:2012dx}. To understand the parton deconfinement at finite temperature, we study in this paper the heavy quark potential in the GZ approach at finite temperature in the frame of resummed perturbation theory. We will focus on the region around the deconfinement temperature $T_c$ ($T\leq 2T_c$) which can be realized in high energy nuclear collisions and where the normal perturbation theories like HTL are not suitable. We calculate firstly the gluon loop, then the gluon propagator in terms of the resummation of gluon and quark loops, and finally the heavy quark potential through constructing the real-time Wilson loop. We summarize in the end.

The Gribov mass parameter $m_G$ is not a free parameter of the theory. It is a dynamical quantity, being determined in a self-consistent way through a gap equation by minimizing the partition function of the system, $\partial\ln Z/\partial m_G^2 = 0$. At lowest level it can be derived from the contribution of a closed loop to the gluon self-energy. In Coulomb gauge it reads~\cite{Zwanziger:2006sc}
\begin{equation}
\label{GZ2}
\int{d^4p\over (2\pi)^4}{1\over (p_0^2+{\bm p}^2){\bm p}^2+m_G^4}={3\over 2N_cg^2},
\end{equation}
where $g(T)$ as a function of temperature is the QCD running coupling constant, and $N_c$ the number of color degrees of freedom. In the imaginary time formalism of finite temperature field theory, the energy integration becomes a Matsubara frequency summation $\int d^4 p/(2\pi)^4=T\sum_n\int d^3{\bm p}/(2\pi)^3$ with $p_0=i\omega_n$ and $\omega_n=2n\pi T$. By evaluating the frequency summation and making the standard $\overline{MS}$ subtraction, the gap equation is simplified as
\begin{equation}
\label{GZ3}
{1\over 4} \ln \left({e\over 2}{\mu^2\over m_G^2}\right)+\int_0^\infty {dx\over u}{1\over e^{m_Gu/T}-1}={3\pi^2\over N_c g^2(T)},
\end{equation}
with $u=\sqrt{x^2+1/x^2}$, where $\mu$ is the renormalization scale that controls the vacuum value of the Gribov mass.

\begin{figure}[htb]
{$$\includegraphics[width=0.35\textwidth]{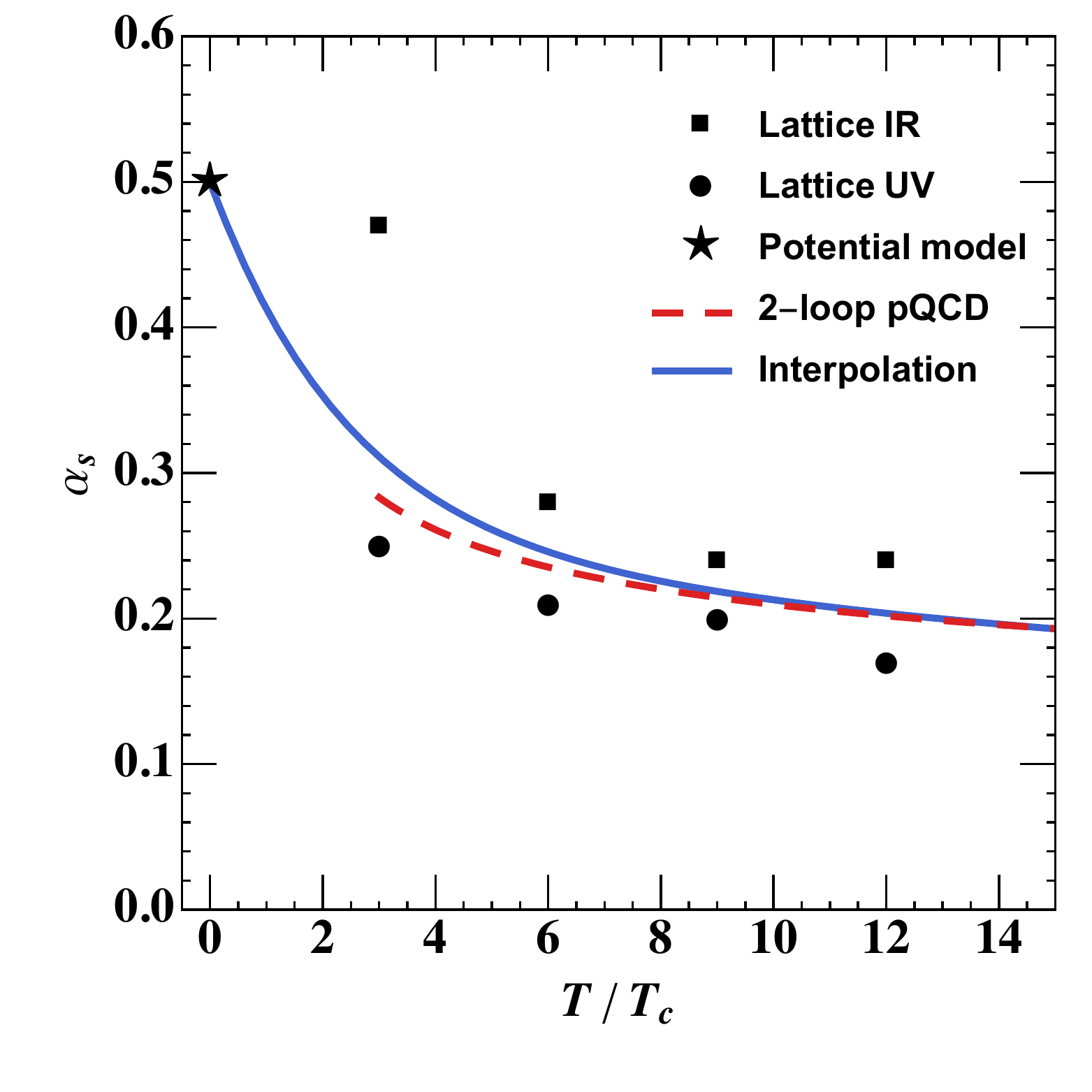}$$
\vspace{-1cm}
\caption{The running coupling constant $\alpha_s(T)$ calculated by lattice IR and UV~\cite{Kaczmarek:2004gv}, perturbative QCD up to two loops~\cite{Kaczmarek:2004gv}, potential models~\cite{Zhao:2020jqu,Lafferty:2019jpr} and interpolation. The deconfinement temperature $T_c$ is chosen to be $170$ MeV~\cite{Lafferty:2019jpr}. }
\label{fig1}}
\end{figure}

The temperature dependence of the Gribov mass is characterized by not only the statistical distribution but also the coupling constant $g^2(T)=4\pi\alpha_s(T)$. At very high temperatures with $T\gtrsim 3T_c$, the IR and UV behaviours of the coupling are obtained through lattice simulations~\cite{Kaczmarek:2004gv}, and the results are qualitatively in consistent with the perturbative QCD calculation up to two loops~\cite{Kaczmarek:2004gv}, as shown in Fig.~\ref{fig1}. In vacuum the coupling can be extracted by fitting the quarkonium spectra via potential model~\cite{Zhao:2020jqu} or fitting the vacuum potential with Cornell potential~\cite{Burnier:2015tda}, both give a coupling $\alpha_s\approx 0.5$. For the gap around $T_c$ where charmonium and bottonium states are expected to be dissociated and we are interested in in this paper, the non-perturbative effect is strong, it is hard to find precise calculation. We then take an interpolation to continuously connect the vacuum and high-temperature values, see the solid line in Fig.~\ref{fig1}.

\begin{figure}[htb]
{$$\includegraphics[width=0.35\textwidth]{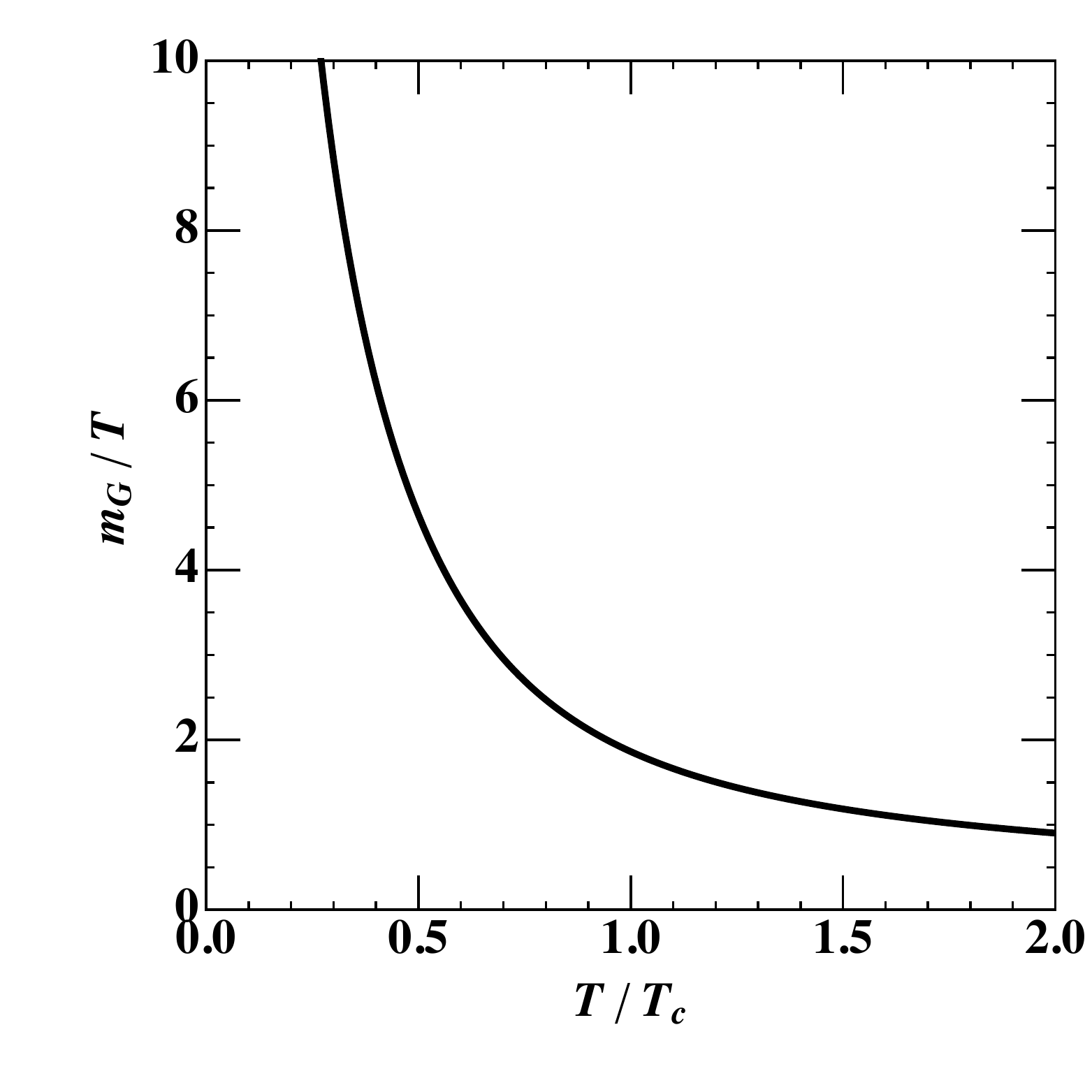}$$
\vspace{-1cm}
\caption{The Gribov mass $m_G(T)/T$.}
\label{fig2}}
\end{figure}

With the running coupling extracted from the interpolation, we solve the gap equation numerically and obtain the Gribov mass $m_G$ as a function of temperature, shown in Fig.~\ref{fig2}. The vacuum value is chosen to be $m_G(0)=0.55$ GeV, corresponding to the renormalization scale $\mu=10.92$ GeV in the gap equation. We will see in the following that this value can reproduce the Cornell potential well. The Gribov mass drops down rapidly in the beginning, when the temperature is below the critical temperature, and then becomes smooth in the deconfinement phase. In the limit of high temperature $T\to \infty$, the solution of the gap equation
\begin{equation}
\label{GZ4}
m_G(T)={N_c\over 2^{3/2}3\pi} g^2(T)T
\end{equation}
approaches to a standard magnetic mass $m\sim g^2(T )T$~\cite{Zwanziger:2006sc,Fukushima:2013xsa}. In the asymptotic free interval with $g^2\to 0$, the Gribov mass disappears.
\vspace{-0.5cm}

To include loop correction to the gluon propagator (\ref{GZ1}), we calculate first the gluon loop. Since the interaction potential between two heavy quarks is only related to the component $D_{00}$ of the gluon propagator $D_{\mu\nu}$, we consider in the following only the component $\Pi_{00}$ of the loop function $\Pi_{\mu\nu}(p)$. Defining $\Pi_G(p)=-\Pi_{00}(p)/(g^2N_c{\bm p}^2)$ and using the free gluon propagator (\ref{GZ1}) in Coulomb gauge, a direct calculation leads to
\begin{eqnarray}
\label{GZ5}
\Pi_G(p) &=& J_1(p)-J_2(p),\nonumber\\
J_1(p) &=& {3p_i p_j \over {\bm p}^2} \int{d^4k\over (2\pi)^4}{T_{ij}({\bm k}) \over ({\bm k}^2+m_G^4/{\bm k}^2-k_0^2)({\bm k}-{\bm p})^2}, \nonumber\\
J_2(p) &=& {1\over {\bm p}^2}\int {d^4k\over (2\pi)^4} {T_{ij}({\bm k})T_{ij}({\bm k}-{\bm p}) \over {\bm k}^2+m_G^4/{\bm k}^2-k_0^2}\nonumber\\
&\times& {{\bm k}^2+m_G^4/{\bm k}^2+k_0^2-k_0p_0 \over ({\bm k}-{\bm p})^2+m_G^4/({\bm k}-{\bm p})^2-(k_0-p_0)^2},
\end{eqnarray}
with $T_{ij}({\bm k})=\delta_{ij}-k_ik_j/{\bm k}^2,\ i,j=1,2,3$. The above two integrals are divergent in vacuum and need to be regularized. Introducing a momentum-cutoff $\Lambda$ and using the renormalization scheme $\overline{MS}$, the divergence can be attracted into the running coupling,
\begin{equation}
\label{GZ6}
{1\over g^2(\Lambda)N_c}= {1\over 48\pi^2}\left(11\log{\Lambda^2\over \Lambda_{\overline{MS}}^2} -{49\over 3}+22\log 2\right),
\end{equation}
where the momentum-cutoff in the scheme $\overline{MS}$ is taken as $\Lambda_{\overline{MS}}\approx m_G/1.62$ with the quark contributions~\cite{Golterman:2012dx}.

$J_2$ can further be separated into three parts,
\begin{eqnarray}
\label{GZ7}
J_2(p) &=& J_{2A}(p)+J_{2B}(p)+J_{2C}(p),\\
J_{2A}(p) &=& {1\over {\bm p}^2}\int {d^4k\over (2\pi)^4} {T_{ij}({\bm k})T_{ij}({\bm k}-{\bm p})\over (k_0-p_0)^2-\epsilon_{k-p}^2},\nonumber\\
J_{2B}(p) &=& {1\over {\bm p}^2}\int {d^4k\over (2\pi)^4} {2\epsilon_k^2T_{ij}({\bm k})T_{ij}({\bm k}-{\bm p})\over (k_0^2-\epsilon_k^2)((k_0-p_0)^2-\epsilon_{k-p}^2)},\nonumber\\
J_{2C}(p) &=& {1\over {\bm p}^2}\int {d^4k\over (2\pi)^4} {-p_0k_0T_{ij}({\bm k})T_{ij}({\bm k}-{\bm p})\over (k_0^2-\epsilon_k^2)((k_0-p_0)^2-\epsilon_{k-p}^2)},\nonumber
\end{eqnarray}
with effective gluon energy $\epsilon_k=\sqrt{{\bm k}^2+m_G^4/{\bm k}^2}$. Considering the fact that a statical potential is calculated at time $t\to\infty$ which corresponds to $p_0\to 0$, $J_{2C}$ disappears in this limit. Taking into account the Matsubara summations over gluon frequency $k_0$,
\begin{eqnarray}
\label{GZ8}
&& T\sum_{k_0}{1\over k_0^2-\epsilon_k^2}=-{1\over 2\epsilon_k}\coth{\epsilon_k\over 2T},\nonumber\\
&& T\sum_{k_0}{1\over (k_0^2-\epsilon_k^2)((k_0-p_0)^2-\epsilon_{k-p}^2)}\nonumber\\
&=&-\sum_{\eta_1,\eta_2=\pm} {\eta_1\eta_2\over 4\epsilon_k\epsilon_{k-p}} {f_B(\eta_2\epsilon_{k-p})-f_B(\eta_1\epsilon_k)\over p_0-\eta_1\epsilon_k+\eta_2\epsilon_{k-p}},
\end{eqnarray}
where $f_B$ is the Bose-Einstein distribution function, $J_1$ and $J_2$ can be expressed as
\begin{eqnarray}
\label{GZ9}
J_1(p) &=& {3p_i p_j\over {\bm p}^2}\int {d^3{\bm k}\over (2\pi)^3} {T_{ij}({\bm k})\over 2\epsilon_k({\bm k}-{\bm p})^2}\coth{\epsilon_k\over 2T}, \nonumber\\
J_{2A}(p) &=& -{1\over {\bm p}^2}\int {d^3{\bm k}\over (2\pi)^3} {T_{ij}({\bm k})T_{ij}({\bm k}+{\bm p})\over 2\epsilon_k}\coth{\epsilon_k\over 2T},\nonumber\\
J_{2B}(p) &=&-{1\over {\bm p}^2}\int {d^3{\bm k}\over (2\pi)^3} \sum_{\eta_1,\eta_2=\pm}{\eta_1\eta_2\epsilon_kT_{ij}({\bm k})T_{ij}({\bm k}-{\bm p})\over 2\epsilon_{k-p}}\nonumber\\
&\times& {f_B(\eta_2\epsilon_{k-p})-f_B(\eta_1\epsilon_k)\over p_0-\eta_1\epsilon_k+\eta_2\epsilon_{k-p}}.
\end{eqnarray}

We now take analytic extension of $p_0\to p_0+i\epsilon$ which leads to the real and imaginary parts of the gluon loop function,
\begin{eqnarray}
\label{GZ10}
\text {Re} \Pi_G(p) &=& J_1(p)-J_{2A}(p)-J_{2B}(p),\nonumber\\
\text {Im} \Pi_G(p) &=& -{\pi\over {\bm p}^2}\int {d^3{\bm k}\over (2\pi)^3} \sum_{\eta_1\eta_2=\pm} {\eta_1\eta_2\epsilon_k\over 2\epsilon_{k-p}}\nonumber\\
&\times& \left[f_B(\eta_2\epsilon_{k-p})-f_B(\eta_1\epsilon_k)\right](1+\cos^2\theta)\nonumber\\
&\times& \delta(p_0-\eta_1\epsilon_k+\eta_2\epsilon_{k-p}),
\end{eqnarray}
where $\theta$ is the angle between the two momentum vectors ${\bm k}$ and ${\bm k}-{\bm p}$, and the $\delta$ function means the energy conservation during the decay process from one gluon to two gluons.

The Gribov region $\Omega$ changes only the path integration of the gauge field, the free quark propagator and in turn the quark loop function are not affected by the Gribov condition and can be found in textbooks~\cite{QFT},
\begin{eqnarray}
\label{GZ11}
\text {Re}\Pi_Q(p) &=& {N_f\over 2N_c {\bm p}^2}\Bigg[{T^2\over 6}+2\int {d^3{\bm k}\over (2\pi)^3} \sum_{\eta_1\eta_2=\pm}\nonumber\\
&\times& {\eta_1\eta_2{\bm k}\cdot({\bm k}-{\bm p})\over |{\bm k}||{\bm k}-{\bm p}|}{f_F(\eta_2|{\bm k}-{\bm p}|)-f_F(\eta_1|{\bm k}|)\over p_0-\eta_1|{\bm k}|+\eta_2|{\bm k}-{\bm p}|}\Bigg],\nonumber\\
\text {Im}\Pi_Q(p) &=& -{\pi N_f\over N_c {\bm p}^2}\int {d^3{\bm k}\over (2\pi)^3} \sum_{\eta_1\eta_2=\pm}{\eta_1\eta_2{\bm k}\cdot({\bm k}-{\bm p})\over |{\bm k}||{\bm k}-{\bm p}|}\nonumber\\
&\times& \left[f_F(\eta_2|{\bm k}-{\bm p}|)-f_F(\eta_1|{\bm k}|)\right]\nonumber\\
&\times& \delta(p_0-\eta_1|{\bm k}|+\eta_2|{\bm k}-{\bm p}|),
\end{eqnarray}
where $f_F$ is the Fermi-Dirac distribution function. While the quark loop $\Pi_Q$ is not explicitly affected by the GZ approach, its renormalization in vacuum is coupled to the Gribov mass $m_G$ through the momentum cutoff  $\Lambda_{\overline{MS}}$.

The total loop function contains both the gluon and quark loops,
\begin{equation}
\label{GZ12}
\Pi(p)=\Pi_G(p)+\Pi_Q(p).
\end{equation}
By summarizing over all gluon and quark loops on a chain~\cite{QFT}, one derives the loop corrected gluon propagator,
\begin{eqnarray}
\label{GZ13}
D_{00}(p) &=& {1\over {\bm p}^2}\left[1-\Pi_{00}(p)/{\bm p}^2+\left(-\Pi_{00}(p)/{\bm p}^2\right)^2+\cdots\right]\nonumber\\
&=& {1\over {\bm p}^2} {1\over 1-g^2 N_c \Pi(p)}.
\end{eqnarray}

We now turn to the calculation of heavy quark potential via the gluon propagator $D$. Aiming to an in-medium potential, we follow the strategy in Ref.~\cite{Laine:2006ns} to construct a real-time Wilson loop which characterizes the propagation of two infinitely heavy quarks. The evolution of the Wilson loop satisfies the Schr\"odinger equation where the potential to the first order of $g^2$ reads~\cite{Laine:2006ns}
\begin{eqnarray}
\label{GZ14}
V_>(t, r) &=& g^2C_F \int {d^3{\bm p} \over (2 \pi)^3}{2-e^{i p_3 r}-e^{-i p_3 r} \over 2}\nonumber\\
&\times& \Bigg\{{1\over {\bm p}^2\left(1-g^2N_c\Pi(0,{\bm p})\right)}+\int {d p_0\over \pi} f_B(p_0)p_0\nonumber\\
&\times& \left(e^{p_0/T}e^{-i p_0 t}-e^{i p_0 t}\right)\Bigg[\left({1\over {\bm p}^2}-{1\over p_0^2}\right) \rho_E(p)\nonumber\\
&-& \left({1\over {\bm p}^2}-{1\over p_3^2}\right) \rho_T (p)\Bigg]\Bigg\},
\end{eqnarray}
with the constant $C_F=(N_c^2-1)/(2N_c)$, where $\rho_E(p)$ and $\rho_T(p)$ are the two spectral functions~\cite{Laine:2006ns}. Using the relation $(e^{i p_0 t}-e^{-i p_0 t})/p_0=2\pi i \delta(p_0)$ and the approximation $f(p_0)\approx T/p_0$ for thermalized gluons, we obtain the statical potential in the limit of $t\to\infty$,
\begin{eqnarray}
\label{GZ15}
V(r)&=& \lim_{t\to \infty} V_>(t, r)=V_R(r)+iV_I(r)\\
V_R(r) &=& -{C_F\over N_c} \int {d^3{\bm p}\over (2\pi)^3}{1\over {\bm p}^2\text {Re}\Pi(0,{\bm p})}\left(1-e^{i p_3 r}\right),\nonumber\\
V_I(r) &=& {C_F\over N_c}T \int {d^3 {\bm p}\over (2 \pi)^3}{{\bm p}^2 F(0,{\bm p})\over |{\bm p}| ({\bm p}^2\text {Re}\Pi(0,{\bm p}))^2}\left(1-e^{i p_3 r}\right)\nonumber
\end{eqnarray}
with
\begin{equation}
\label{GZ16}
F(p) = 2{|{\bm p}|\over p_0}\text {Im}\Pi(p).
\end{equation}

\begin{figure}[htb]
{$$\includegraphics[width=0.35\textwidth]{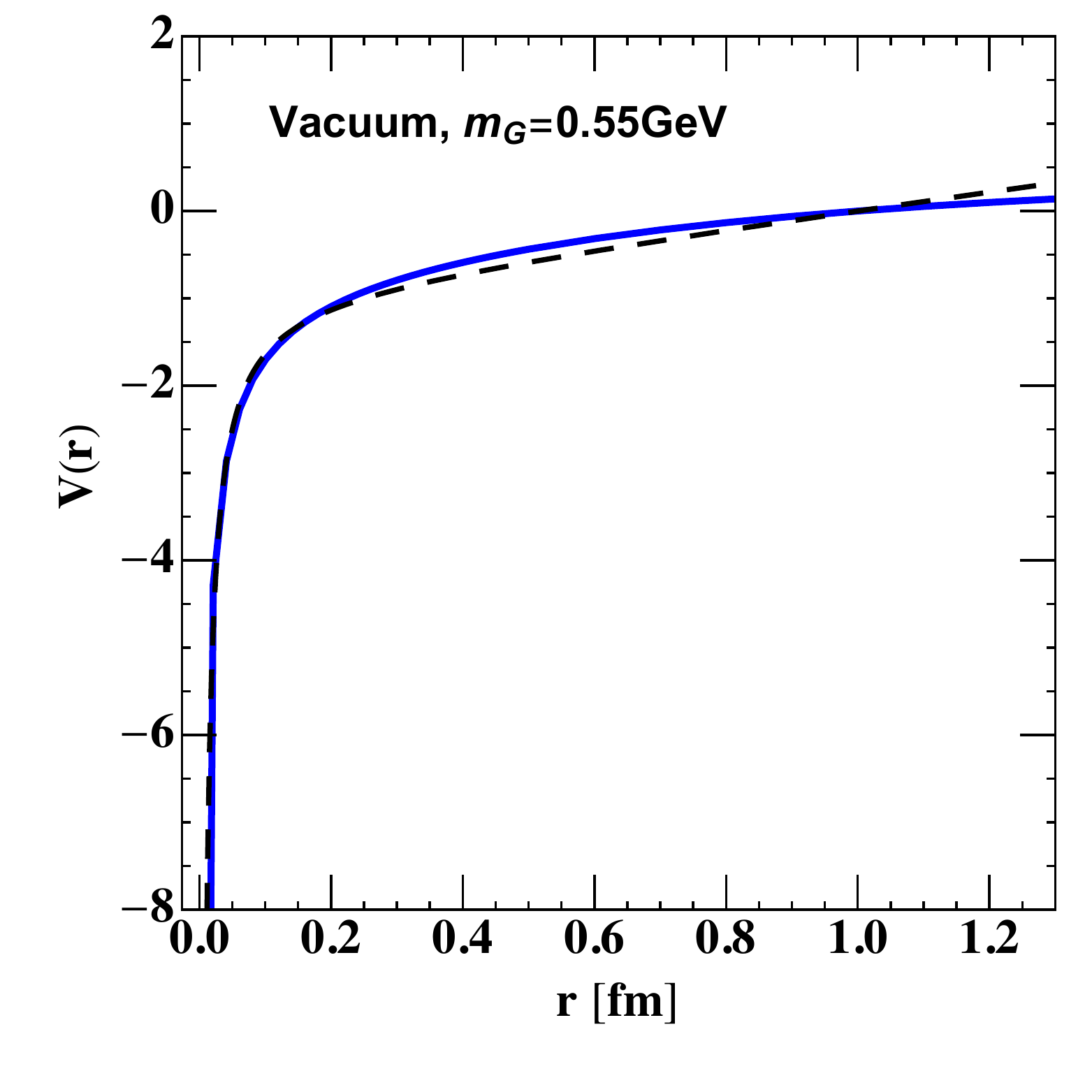}$$
	\vspace{-1cm}
\caption{The vacuum potential $V(r)$ scaled by the condition $V=0$ at $r=1$ fm. The dashed and solid lines are calculated via Cornell potential and Gribov-Zwanziger approach. }
\label{fig3}}
\end{figure}

In vacuum, there is no Landau damping, the imaginary parts of the loop function, propagator and potential disappear automatically, and the potential is reduced to
\begin{equation}
\label{GZ17}
V(r) = C_F g^2\int {d^3{\bm p} \over (2\pi)^3}D_{00}(0,{\bm p})\left(1-e^{i{\bm p}\cdot {\bm r}}\right).
\end{equation}
Taking the Gribov mass $m_G=0.55$ GeV, which is approximately the value used in Ref.~\cite{Golterman:2012dx}, the one-loop corrected potential can reproduce very well the Cornell potential $V(r)=-\alpha/r+\sigma r$ with $\alpha=0.4105$ and $\sigma=0.2~\text{GeV}^2$~\cite{Zhao:2020jqu}, see the comparison in Fig.~\ref{fig3}.
\begin{figure}[htb]
{$$\includegraphics[width=0.35\textwidth]{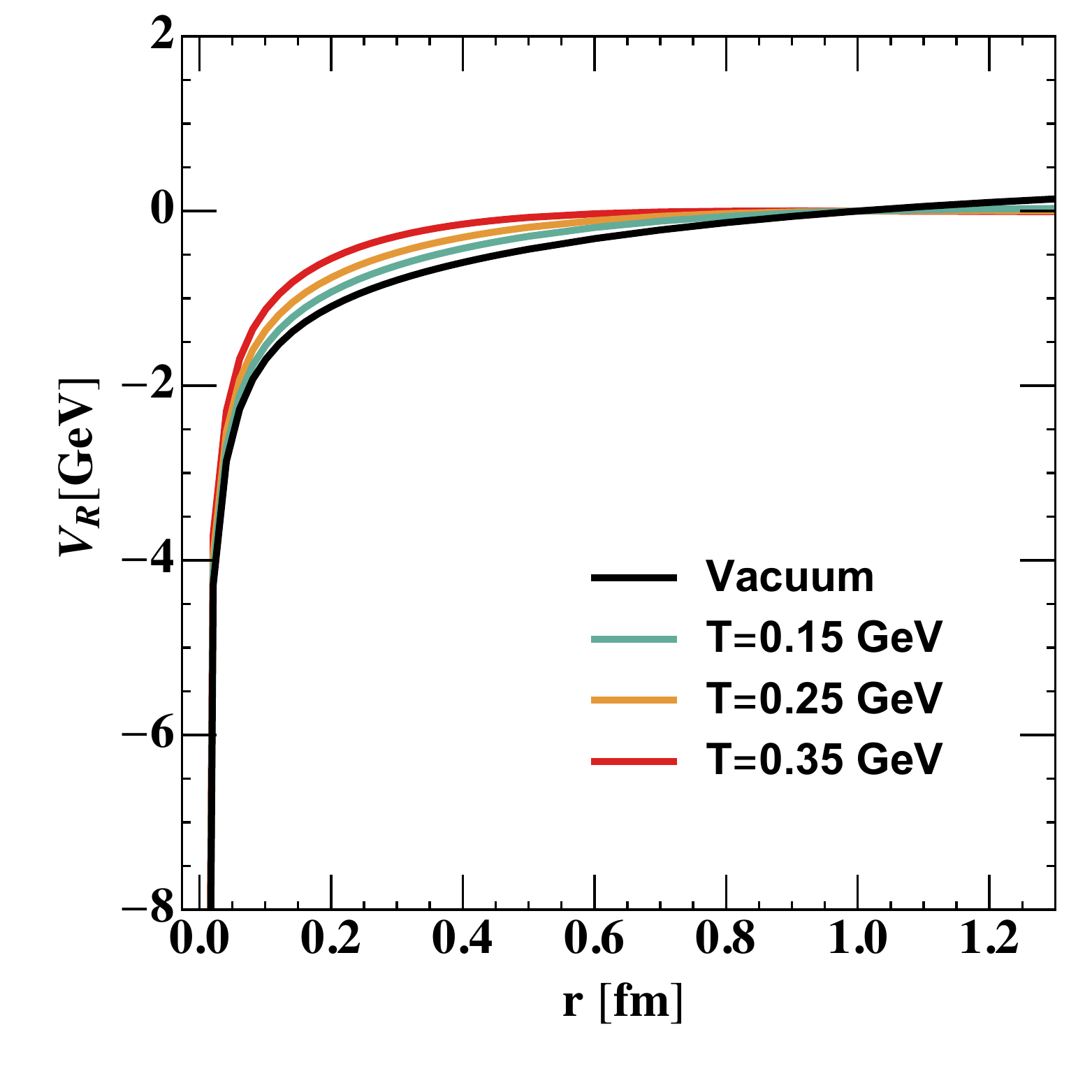}$$
\vspace{-1cm}
\caption{The real potential $V_R(r)$ at finite temperature, again scaled by the condition $V_R=0$ at $r=$ 1 fm. }
\label{fig4}}
\end{figure}
\begin{figure}[htb]
{$$\includegraphics[width=0.35\textwidth]{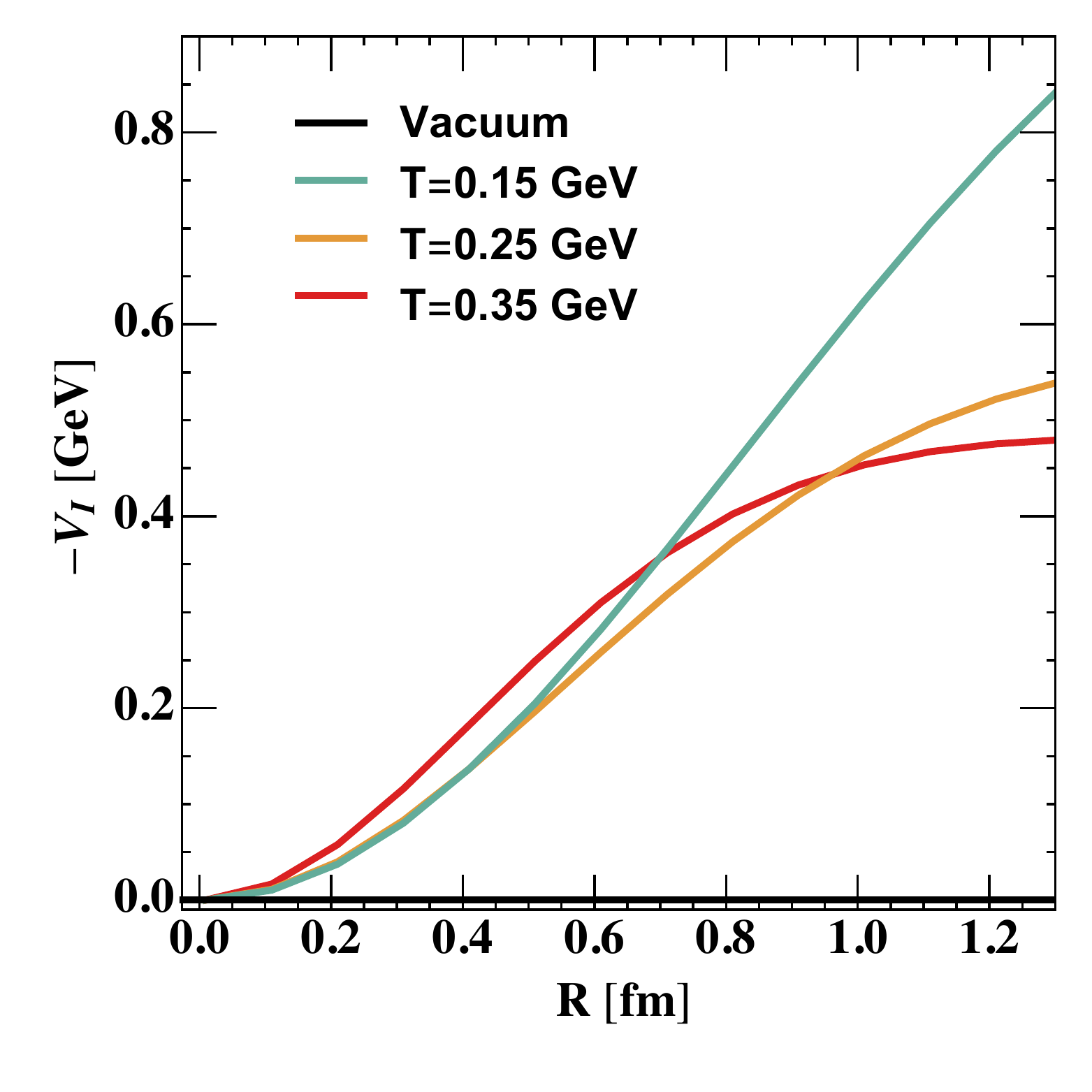}$$
\vspace{-1cm}
\caption{The imaginary potential $V_I(r)$ at finite temperature. }
\label{fig5}}
\end{figure}
\begin{figure}[htb]
{$$\includegraphics[width=0.35\textwidth]{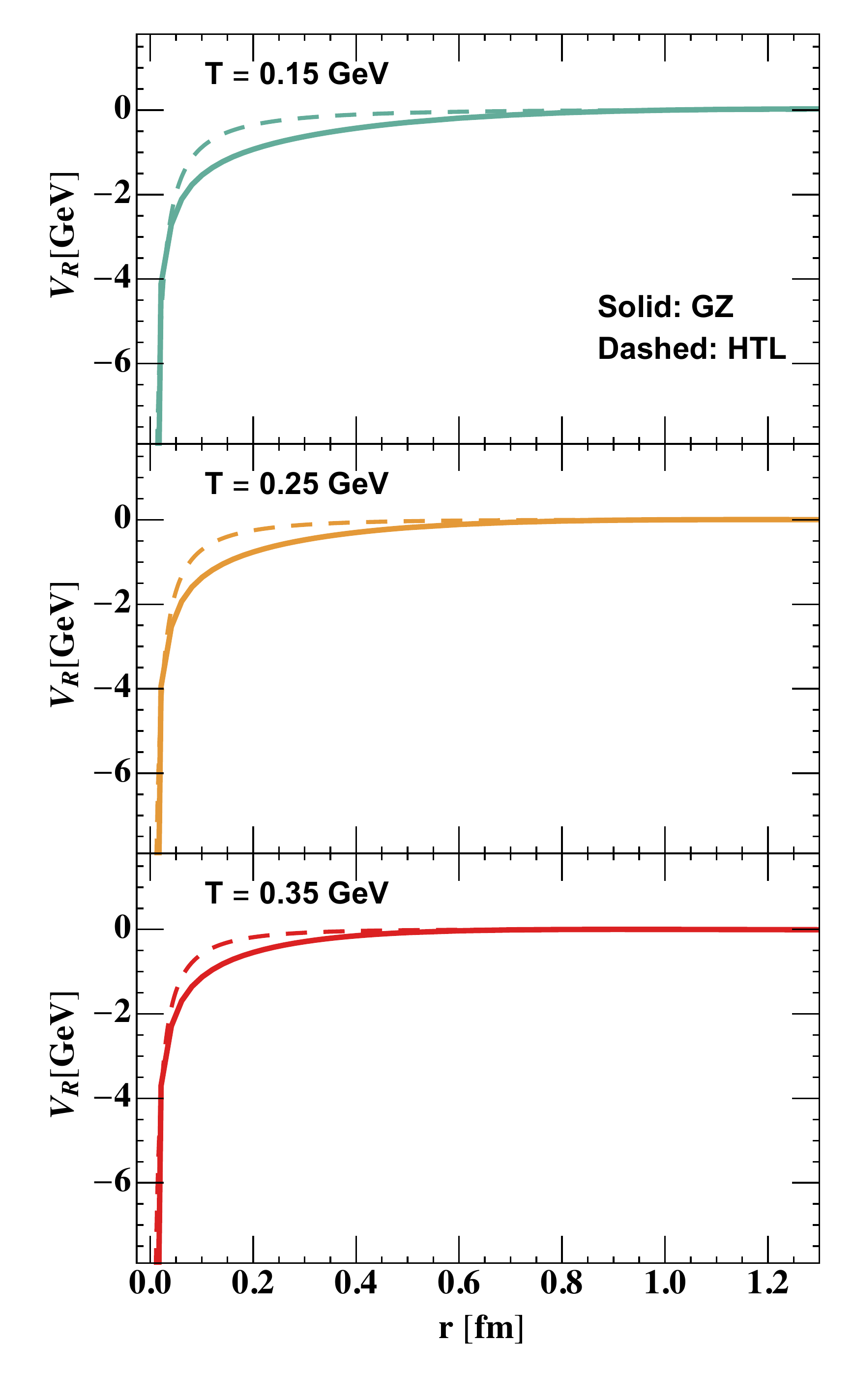}$$
\vspace{-1cm}
\caption{The comparison of the real potentials $V_R(r)$ calculated via Gribov-Zwanziger approach (solid lines) and HTL approach (dashed lines). }
\label{fig6}}
\end{figure}
\begin{figure}[htb]
{$$\includegraphics[width=0.35\textwidth]{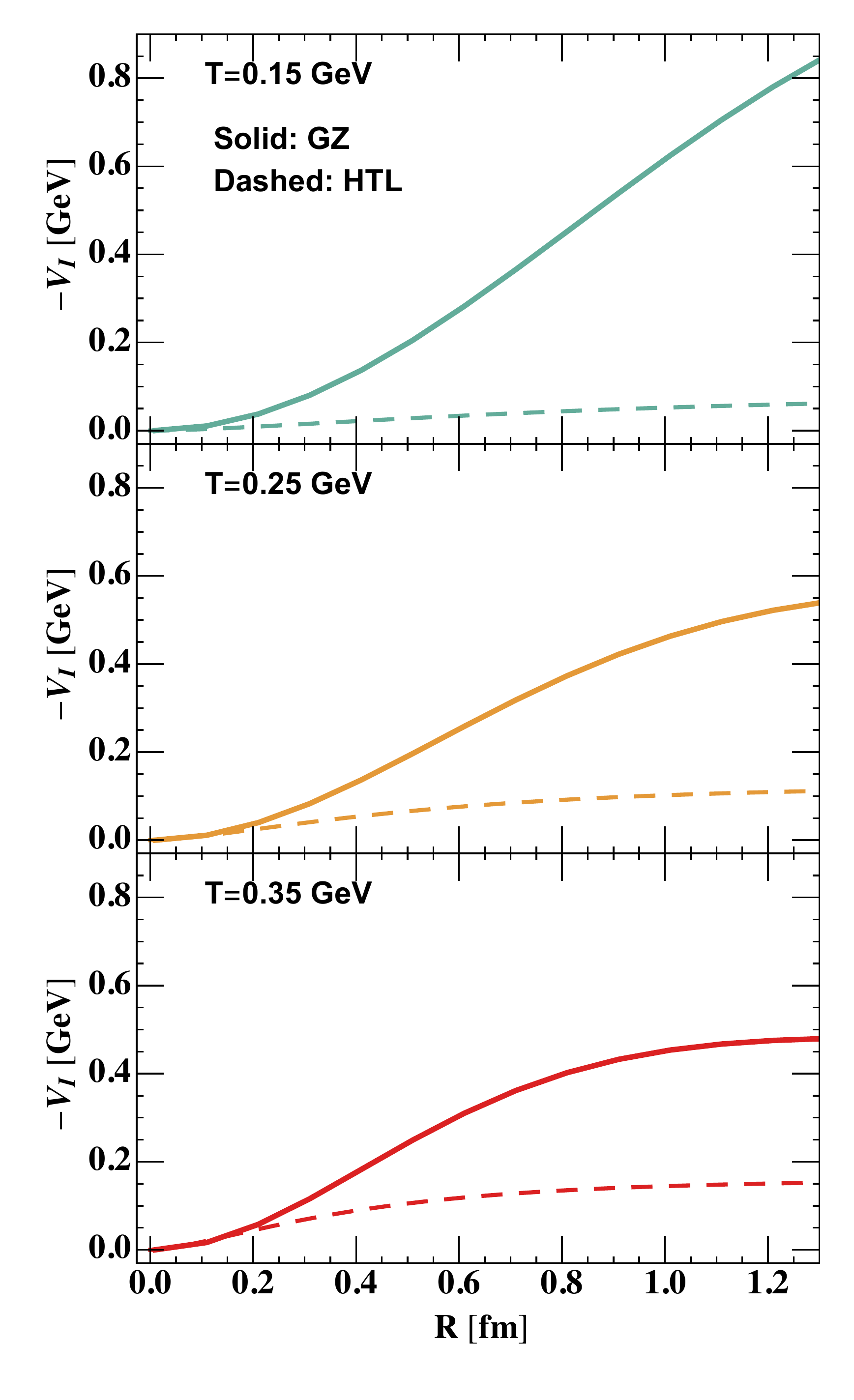}$$
\vspace{-1cm}
\caption{The comparison of the imaginary potentials $V_I(r)$ calculated via Gribov-Zwanziger approach (solid lines) and HTL approach (dashed lines). }
\label{fig7}}
\end{figure}

The heavy quark potential in medium contains real and imaginary parts, shown in Figs.~\ref{fig4} and \ref{fig5} at several temperatures. The real part $V_R(r)$ is, as usually discussed in literatures~\cite{Satz:2005hx}, controlled by color screening. In vacuum, the potential increases linearly with the distance and is never saturated, which means the parton confinement. In hot medium, the potential becomes saturated at a finite distance (Debye screening length $r_D$), due to the color screening. In the saturated region, the interaction force ${\bf F}=-{\bm \nabla}V=0$ indicates the vanishing color interaction at the distance $r>r_D$. With increasing temperature, the screening length decreases from infinity in vacuum to about 0.8, 0.6 and 0.4 fm at $T=0.15$, $0.25$ and $0.35$ GeV. Considering that $r_D=0.4$ fm is already less than the $J/\psi$ radius $r_{J/\psi}\sim 0.5$ fm, charmonia are expected to be dissociated in a fireball with temperature $0.25<T<0.35$ GeV.

For a dynamically evaluating system created in high energy nuclear collisions, quarkonia are dissociated by not only the color screening, but also the Landau damping at finite temperature which leads to a decay width characterized by the imaginary part $V_I(r)$ in potential models. Our calculation with GZ approach shows a strong imaginary part of the potential. Especially at the critical distance of charmonium radius $r\sim 0.5$ fm, the strengths of the real and imaginary parts are at the same order, see Figs.~\ref{fig4} and \ref{fig5}.

It is clear that the one-loop corrected potential should approach to the HTL result in high temperature limit~\cite{Laine:2006ns}, when the Gribov mass $m_G(T)$ goes to zero,
\begin{eqnarray}
\label{GZ18}
V(r) &=& -{C_Fg^2\over 4\pi}\left[\left(m_D+{e^{-m_Dr}\over r} \right)+iT\phi(m_Dr)\right],\nonumber\\
\phi(x) &=& 2\int_0^\infty dz{z \over (z^2+1)^2}\left(1-{\sin(zx) \over zx}\right),
\end{eqnarray}
where $m_D=gT\sqrt{N_c/3+N_f/6}$ is the Debye mass in HTL approximation. From the comparison between the two real potentials (\ref{GZ15}) with GZ approach and (\ref{GZ18}) with HTL approach, shown in Fig.~\ref{fig6}, while the two calculations are close to each other at high temperature, indicating the almost same color screening strengths in the two approaches, the HTL approach fails to go back to a confinement potential at low temperature, according to its definition. As for the imaginary part, shown in Fig.\ref{fig7}, the strength of the one with GZ approach is much larger than the one with HTL approach. The physics is the Gribov mass which behaviors like a magnetic mass $m_G\sim g^2T$ at high temperature. The still large difference between the two imaginary parts at temperature $T=0.35$ GeV $\sim 2 T_c$ comes from two reasons: the temperature is not high enough to satisfy the HTL condition $T\gg |{\bm p}|$, and the Gribov mass is still large $m_G\sim 0.3$ GeV. The slowly dropping down of the Gribov mass at temperature $T\gtrsim T_c$ is probably due to the gap equation at the lowest level. The loop correction to the gap equation should be considered in the future.

Confinement in vacuum and deconfinement in hot medium are non-perturbative problems in QCD, but they cab be described by the gluon propagator at lowest level in the Gribov-Zwanziger approach. This leads to an alternative way to perturbatively calculate the heavy quark potential through the gluon propagator. We calculated in this paper the loop-corrected heavy quark potential via the resummation of gluon and quark loops in the strong coupling region around the deconfinement phase transition temperature $T_c$. The loop correction makes the potential complex, the saturation of the real part is controlled by the color screening, and the imaginary part is characterized by the Landau damping in hot medium. In comparison with the HTL approach of QCD, the high temperature behavior of the real part is almost the same in the two approaches, while the Gribov mass results in a much stronger imaginary part due to effectively including the magnetic interaction in the GZ approach. For the distance around the typical quarkonium size, the strength of the imaginary part is comparable with the real part. This means that, different from the previous quarkonium dissociation picture by color screening, Landau damping plays also an important role.

{\bf Acknowledgement}: The work is supported by the Guangdong Major Project of Basic and Applied Basic Research No. 2020B0301030008 and the NSFC grants Nos. 11890712 and 12075129.


\end{document}